\documentclass[notitlepage,preprint]{article}
\usepackage[utf8]{inputenc} 
\usepackage[colorlinks=true,urlcolor=blue,linkcolor=blue, citecolor=blue]{hyperref}
\usepackage{graphicx}
\usepackage[a4paper,left=3cm, right=3cm, top=2cm, bottom=3cm]{geometry}
\usepackage{authblk}

\usepackage{amsthm}
\usepackage{amsmath}

\title{The Latin American Giant Observatory: a successful collaboration in Latin America based on Cosmic Rays and computer science domains}

\author[1]{H. Asorey\thanks{\href{mailto://asoreyh@cab.cnea.gov.ar}{asoreyh@cab.cnea.gov.ar}}}
\author[2]{R. Mayo-García\thanks{\href{mailto://rafael.mayo@ciemat.es}{rafael.mayo@ciemat.es}}}
\author[3,4]{L. A. Núñez\thanks{\href{mailto://lnunez@uis.edu.co}{lnunez@uis.edu.co}}}
\author[2]{M. Rodríguez-Pascual\thanks{\href{mailto://manuel.rodriguez@ciemat.es}{manuel.rodriguez@ciemat.es}}}
\author[2]{A. J. Rubio Montero\thanks{\href{mailto://antonio.rubio@ciemat.es}{antonio.rubio@ciemat.es}}}
\author[3]{M. Suarez-Durán\thanks{\href{mailto://mauricio.suarez@correo.uis.edu.co}{mauricio.suarez@correo.uis.edu.co}}}
\author[4]{L. A. Torres-Niño\thanks{\href{mailto://alejandro.torres@correo.uis.edu.co}{alejandro.torres@correo.uis.edu.co}}}
\author[5]{for the LAGO Collaboration\thanks{\href{mailto://lago@lagoproject.org}{lago@lagoproject.org}}}

\affil[1]{Laboratorio Detección de Partículas y Radiación, Instituto Balseiro \& Centro Atómico Bariloche, San Carlos de Bariloche, Argentina}
\affil[2]{División de Tecnologías de la Información y las Comunicaciones, Centro de Investigaciones Energéticas Medioambientales y Tecnológicas, Madrid, Spain}
\affil[3]{Escuela de Física, Universidad Industrial de Santander, Bucaramanga, Colombia}
\affil[4]{Centro de Supercomputación y Cálculo Científico, Universidad Industrial de Santander, Bucaramanga, Colombia}
\affil[5]{\href{http://lagoproject.org}{lagoproject.org}, see the full list of members and institutions at \href{http://lagoproject.org/collab.html}{lagoproject.org/collab.html}.}
\date{\today}

\begin{document}
\maketitle
\begin{abstract}
In this work the strategy of the Latin American Giant Observatory (LAGO) to build a Latin American collaboration is presented. Installing Cosmic Rays detectors settled all around the Continent, from Mexico to the Antarctica, this collaboration is forming a community that embraces both high energy physicist and computer scientists. This is so because the data that are measured must be analytical processed and due to the fact that \textit{a priori} and \textit{a posteriori} simulations representing the effects of the radiation must be performed. To perform the calculi, customized codes have been implemented by the collaboration. With regard to the huge amount of data emerging from this network of sensors and from the computational simulations performed in a diversity of computing architectures and e-infrastructures, an effort is being carried out to catalog and preserve a vast amount of data produced by the water-Cherenkov Detector network and the complete LAGO simulation workflow that characterize each site. Metadata, Permanent Identifiers and the facilities from the LAGO Data Repository are described in this work jointly with the simulation codes used. These initiatives allow researchers to produce and find data and to directly use them in a code running by means of a Science Gateway that provides access to different clusters, Grid and Cloud infrastructures worldwide.

\end{abstract}

\section{Introduction}
The LAGO (Latin American Giant Observatory) project is an extended Astroparticle Observatory at global scale. It is mainly oriented to basic research on three branches of Astroparticle physics: the Extreme Universe, Space Weather phenomena, and Atmospheric Radiation at ground level.

The LAGO detection network consists in single or small arrays of particle detectors at ground level, spanning over different sites located at significantly different latitudes (currently from Mexico up to the Antarctic region) and different altitudes (from sea level up to more than $5,000$ meters over sea level), covering a huge range of geomagnetic rigidity cut--offs and atmospheric absorption/reaction levels.

The LAGO Project is operated by the LAGO Collaboration, a non-centralized and distributed collaborative network of more than $80$ scientist from more than $25$ institutions of $10$ Latin American (currently Argentina, Bolivia, Brazil, Colombia, Ecuador, Mexico, Peru and Venezuela) and European (Spain) countries.

Also, detectors installed in various universities are used as a tool to teach students about particle and astroparticle physics, in particular by leading them to the measurement of the muon decay.

Today, LAGO is a network of ground--based Water Cherenkov Detector (WCDs) whose data are of interest for two different scientific communities:
\begin{itemize}
\item Gamma Astronomy Community: The LAGO WCDs installed at high altitude sites are sensitive to detect the effects of GRB. A significant number of LAGO detectors are over $3,000$ m a.s.l. and three of them are above $4,300$ m.
\item Space Weather: The LAGO energy range covers a plethora of phenomena related to low-energy cosmic rays physics, from solar activity to space weather phenomena. Nowadays, the study of such phenomena is crucial to be studied because levels of radiation in the atmosphere and near-Earth space environment may be established.
\end{itemize}

Since the LAGO data analysis needs to take into account the influence of atmospheric effects, such as the pressure or the air temperature, on the flux of particles at the detector level, each LAGO WCD is equipped with several environmental sensors. These measurements represent an opportunity to provide environmental information to other communities, like ecologists studying the high altitude environments to correlate for possible climate change and global warming effects. Additionally, since the data in the LAGO repositories is open and freely accessible, it is used as motivation to train the general public -mainly the secondary school teachers and students- in statistical data analysis and related techniques, and raising the awareness of the general public in global warming and climate change impact in everyday life. This important citizen science initiative is one of the main objectives of the LAGO collaboration and is implemented in the so-called LAGO-CS (Citizen Science) program.  

To properly analyze and exploit the measured data, computational capabilities must be available to the collaboration. This topic is being provisioned on a two-fold basis: customized codes that replicate the phenomena happened in the LAGO sites/detectors and data and metadata functionalities to be curated and managed.

With respect to the former, the CORSIKA code \cite{Corsika} used by LAGO covers three different energy ranges, which are different to the ones usually studied in other consortia such as the Pierre Auger Observatory\cite{PierreAuger}.

On the other side, new challenges are emerging around the complex discovery environments in Data Centered Science, where trustworthiness and reproducibility of data are key requirements for new scientific findings. Experimental protocols are central part of any research and they are aimed to ensure the replicability of the results. 

An increasingly frequent scenario is a researcher examining online the existing bibliography on a particular area. He/She finds several publications based on significant amount of data, registered, simulated or both and simultaneously is automatically redirected to the data used to produce them. Later on, the researcher access to the corresponding application employed to generate simulated data as well as the recorded raw/processed data. The new data, synthetic, measured or both (and the new paper if any) are stored on the Data Infrastructure and can be easily found making possible to start the cycle again.

This open access to data and/or applications will help to make reproducible the increasing complex workflow for producing scientific knowledge today.

\section{Some hints about the LAGO related Physics}

Nowadays, the LAGO Project is an extended Astroparticle Observatory at a continental scale, mainly oriented towards developing astroparticle physics at Latin America and doing basic research in three areas: search for the high energy component of Gamma-ray bursts at high altitude sites, Space Weather phenomena, and Background Radiation at ground level \cite{4}. The LAGO detection network consists in particle detectors deployed at ground level, spanning over different sites located at significantly different latitudes (currently planned from Mexico down the Antarctic region) and different altitudes (from sea level up to more than $5,000$ meters over sea level), covering a large range of geomagnetic rigidity cut-offs and atmospheric absorption/reaction levels \cite{5}.

The current network of detectors is operated by the LAGO Collaboration, a non-centralized and distributed collaborative network of more than $80$ scientist from institutions of nine Latin American countries (currently Argentina, Bolivia, Brazil, Colombia, Ecuador, Guatemala, Mexico, Peru and Venezuela) and Spain. Due to its proved reliability, high detection efficiency to all components present in atmospheric extensive showers, and low cost, water Cherenkov detectors (WCD) are currently used at LAGO sites \cite{3}.

The LAGO simulation chain consists of:
\begin{itemize}
\item Dynamic directional rigidity cut-off at each site R(Lat; Lon; $\theta$; $\phi$; $t$)
\item Primary flux at the top of the atmosphere, i.e. CORSIKA simulations for each site ($\varphi$; $\lambda$; $h$)
\begin{itemize}
\item Measured spectra for all nuclei $1 \leq Z_p \leq 26$, $1 \leq A_p \leq 56$
\item $\left[R(\theta; \varphi) \times Z_p\right] \leq (E_p/GeV) \leq 10^6$, $0^{\circ} \leq \theta \leq 90^{\circ}$
\item Integrated primary flux:  $10^7 - 10^8$ hour$^{-1}$ m$^{-2}$ ($\geq 5$ hours at each site)
\end{itemize}
\item Secondary flux at detector level
\item Detector response:
\begin{itemize}
\item Fast and Simple LAGOFast detector simulation
\item Detailed GEANT4 model
\end{itemize}
\end{itemize}

\section{LAGO simulations}
The main code used in LAGO is CORSIKA. CORSIKA (COsmic Ray SImulations for KAscade ) is a program for detailed simulation of extensive air showers initiated by high energy cosmic ray particles. Protons, light nuclei up to iron, photons, and many other particles may be treated as primaries.

The particles are tracked through the atmosphere until they undergo reactions with the air nuclei or (in the case of unstable secondaries) decay. The hadronic interactions at high energies may be described by several reaction models alternatively: The VENUS, QGSJET, and DPMJET models are based on the Gribov-Regge theory, while SIBYLL is a minijet model. The neXus model extends far above a simple combination of QGSJET and VENUS routines. The most recent EPOS model is based on the neXus framework but with important improvements concerning hard interactions and nuclear and high-density effect. HDPM is inspired by findings of the Dual Parton Model and tries to reproduce relevant kinematical distributions being measured at colliders.

LAGO aims to study cosmic rays in the energy range $10$GeV--$100$TeV. In this energy range there emerges phenomena related to the physics of low-energy cosmic rays, and also to solar activity and space weather environment. Nowadays it is crucial to study of these effects because it may establish levels of radiation in the atmosphere and near-Earth space environment.

To do so, the LAGO collaboration has implemented three specific and customized versions of CORSIKA that are to be executed on either local clusters or distributed environments such as Grid and Cloud.

In addition to CORSIKA, LAGO uses intensively other important codes: MAGNETOCOSMIC\cite{Desorgher2004}, GEANT4\cite{AgostinelliEtal2003}, ROOT\cite{KumarTripathi2008} and specific self-designed statistical codes for data analysis, focusing most of the collaboration activities (research \& outreach) on a data repository.

A fully dedicated Virtual Organization, called \textit{lagoproject} is already integrated into the European Grid Infrastructure (EGI)\footnote{http://www.egi.eu} activities. The Grid implementation of CORSIKA was deployed in two 'flavors' being able to run by using GridWay Metascheduler\footnote{http://www.gridway.org/doku.php} \cite{HuedoEtal2001} or with a second approach through a Catania Science Gateway interface\cite{BarberaFargettaRotondo2011}. For the former, massive calculations can be executed via the Montera \cite{Montera}, the GWpilot \cite{GWpilot} or the GWcloud \cite{Gwcloud} frameworks

In the Science Gateway approach a user can seamlessly run a code on different infrastructures by accessing a unique entry point with an identity provision. He/she only has to upload the input data or use a PID to reference it and click on the run icon. The final result will be retrieved whenever the job will be ended. The underlying infrastructure is absolutely transparent to the user and the system decides on which sites and computing platform the code is performed.

\section{The LAGOData e-infrastructure}

Typically, each detector generates $150$ GB of data per month and the entire collaboration generates $1.5$ TB/month. The LAGO dataset not only refers to data measured by WCD detectors but also to data generated by simulation of cosmic rays phenomena in the aforementioned energy range. The CORSIKA particle flux simulations carried out generate $10$GB/site and these synthetic data are also preserved in the data repository. 

The low energy limit depends on the geomagnetic coordinates of the site, while the high energy limit is determined by the collection area at each site and is limited by statistics as the flux becomes lower and lower at higher energies (in general, the cosmic rays flux decreases by a factor of $1000$ for an increase of $10$ in the cosmic ray energy). This raw data collected by the LAGO detectors are shared trough LAGOData\cite{TorresEtAl2011}, a platform  conceived to promote data curation and sharing among LAGO collaborators, which is part of a more ambitious project, LAGOVirtual\cite{CamachoEtal2009} a working environment which ensure access to the data recorded in all LAGO Sites and facilitate the analysis of such data. 

\subsection{Data curation through DSpace}
Dspace is an open source software that enables sharing of many types of content, it is generally used for institutional repositories, providing basic functionality for saving, storing, and retrieving of digital content. DSpace was adopted for the LagoDATA repository, because it hosts Dublin Core metadata with a straightforward adaptability for non-native metadata schemes. It also supports two important interoperability protocols: OAI-PMH (Open Archive Initiatives Protocol for Metadata Harvesting\footnote{http://www.openarchives.org/pmh/} ) and SWORD (Simple Webservice Offering Repository Deposit\footnote{http://swordapp.org/}). The OAI-PMH protocol at the LAGO repository allows the CHAIN-REDS Knowledge Base search engine to navigate into LAGO curated data. 

It was important to overcome one of the most important DSpace limitations, i.e. its inability to upload/download multiple records. Dspace offers the possibility to upload the corresponding metadata through a command line option via an \textit{import tool} by using \textit{simple archive format} and including it in a separate way. A script to ingest data profiting from the above mentioned DSpace capability has been developed as well. With this scheme, it is worth mentioning that part of the metadata is the PID  

\subsection{The LAGOData metadata}
The Dublin Core metadata element set is a standard for cross-domain information resource description, it is elaborated and sponsored by DCMI (Dublin Core Metadata Initiative\footnote{http://dublincore.org/}), the implementation of which makes use of XML. Dublin Core is Resource Description Framework based\footnote{http://www.w3.org/RDF/} and comprises fifteen metadata elements: Title, Subject, Description, Source, Language, Relation, Coverage, Creator, Publisher, Contributor, Rights, Date, Type, Format, and Identifier.  Despite this functionality is mostly centered on the Dublin Core metadata scheme, the additional non-native metadata can be configured as custom fields which are also stored, searched and displayed as the native ones.

The datasets are classified into three different types with their corresponding associated metadata: WCD, simulated, and calibration. Thus:
\begin{itemize}
\item WCD metadata scheme is: \textsf{data} corresponds to the version/type of the Digit/Analog electronic board; \textsf{site} contains the \textit{name}, \textit{latitud},  \textit{longitude} and \textit{heigth} of the detector; \textsf{voltage}, \textsf{level} and \textsf{sensor};
\item simulation metadata uses: \textsf{primary} described by the CORSIKA input file DATXXXX.dbase; \textsf{site} with the \textit{latitud},  \textit{longitude} and \textit{heigth} of the ground point;\textsf{libraries} indicating which are the included CORSIKA libraries; \textsf{computation} describing the computational environment by unix command \texttt{uname -a}, \texttt{lsb\_release -a}, \texttt{free} and \texttt{gcc -v}; 
\item calibration data refers to the calibration parameters used in the LAGO site.
\end{itemize}

\subsection{PID and LAGOData}
The main interface to register and manage PID services for European Research Communities is EPIC (European Persistent Identifiers Consortium\footnote{http://pidconsortium.eu}) which is based on the Handle System
\footnote{The Handle System (http://www.ietf.org/rfc/rfc3650.txt). This provides efficient, extensible, and secure resolution mechanism for unique and persistent identifiers of digital objects. The Handle System includes an open set of protocols (http://hdl.handle.net/4263537/4086), a namespace (http://hdl.handle.net/4263537/4068), and a reference implementation (http://handle.net/download.html) of the protocols.}
for the allocation and resolution of persistent identifiers. 
There are several compatible 'flavors' of PIDs. The most common is DOI PIDs\footnote{The Digital Object Identifier (DOI) System (http://www.doi.org/), a service operated by the International DOI Foundation (IDF), which provides a technical and social infrastructure for the registration and use of persistent interoperable identifiers on digital networks.}. DOI PIDs are more frequently used for publications while the EPIC PIDs cover a wider range of Digital objects. 

The GRNET PID service\footnote{http://epic.grnet.gr} enables the allocation, management and resolution of PIDs and has been employed to ensure the data persistence and reproducibility of the experiments. It supports the use of part identifiers as they are provided by the Handle system. Part identifiers can compute an unlimited number of handles on the fly, without requiring registering each separately. 

\subsection{SWORD and LAGOData}
The SWORD (Simple Web-service Offering Repository Deposit) \cite{LewisDeCastroJones2012}, based on the Atom Publishing Protocol (AtomPub), was first developed to standardize a deposit interface to digital repositories. Presently, it further extend the limited capabilities of AtomPub by supporting the whole deposit lifecycle, i.e. deposit, update, replace, and delete resources. Many interfaces of laboratory equipment allows automatic capture of results in an information system and SWORD permits to upload data directly into a repository, without human intervention, tagging as metadata how data was collected and the conditions in which were collected.    

\subsection{The DART challenge in LAGO}
The Data Accessibility, Reproducibility and Trustworthiness (DART) initiative was launched by CHAIN-REDS\footnote{http://www.chain-project.eu} (Coordination and Harmonisation of Advanced e-infrastructure for Research and Education Data Sharing), an European Commission co-funded project focused on promoting and supporting technological and scientific collaboration across different communities in various continents. This initiative provided a set of interrelated tools and services, based on worldwide adopted standards, which made possible to easily/seamlessly access datasets, data/documents repositories and the applications that could generate and/or make use of them.

\textit{Trustworthiness} can be associated to data curation, particularly on the quality of the metadata describing the experimental protocol and data provenance, while \textit{reproducibility} and  \textit{replicability} is closely connected to the accessibility to  data sources and the possibility to manipulate/analyze data contained in them.

CHAIN-REDS approach to data trustworthiness and reproducibility is based on the integration of computational resources and services, with three main cornerstones:
\begin{enumerate}
\item adoption of standards for data discoverability, provenance and recoverability: OAI-PMH\footnote{http://www.openarchives.org/pmh/} for metadata retrieval, Dublin Core\footnote{http://dublincore.org}, as metadata schema, SPARQL\footnote{http://www.w3.org/2001/sw/wiki/SPARQL} for semantic web search and XML\footnote{http://www.w3.org/XML/} as potential standard for the interchange of data;
\item enablement of datasets authorships and user authentication with the corresponding assignments of specific roles on data services, which can be implemented by two strategies:  
\begin{itemize}
\item assignment of PIDs to name data in a unique and timeless manner, ensuring that future changes on URIs or internal organization of databases will be transparent to the user
\item implementation of federated identity provision, a secure, flexible and portable  mechanism to access e-infrastructures worldwide, based on agreements and standards. 
\end{itemize}

\item access to a plethora of computing power to analyze the retrieved data or to contrast them to simulations through an intuitive web-interface. Over the last years, Science Gateways have risen as an ideal tool to allow scientists across the world to seamlessly access different ICT-based infrastructures for research activities to support their day-by-day work and do better (and faster) research.  
\end{enumerate}

In other words: user identification; execution of distributed applications with a simple web interface; usage of Open Access Document Repositories and Data Repositories; and reproducibility of the experiments conform the DART challenge, where the whole research cycle is covered and can be seamlessly performed by non-expert users, hiding complex processes under simple interfaces and minimizing the need for learning new tools \cite{CSE, DART}.

\section{Conclusion}
The LAGO collaboration is working on studying cosmic rays in the energy range $10$GeV--$100$TeV, which complements other astroparticle initiatives. In this energy range there emerges phenomena related to the physics of low-energy cosmic rays, and also to solar activity and space weather environment. Nowadays it is crucial to study these effects because they may establish levels of radiation in the atmosphere and near-Earth space environment. Thus the data repository (and the network of data repositories) will be of interest not only for LAGO and even the cosmic ray community, but useful for the solar physics and space climatology communities.

Such a research activity is being carried out by fostering collaboration in Latin America using experimental sites allocated along the continent as well as Cluster, Cloud and Grid Computing resources geographically distributed too.

\section*{Acknowledgment}
The LAGO Collaboration is very thankful to the Pierre Auger Collaboration for its continuous support


\begin{thebibliography}{10}

\bibitem{Corsika}
D.~Heck \textit{et al.}, \emph{CORSIKA: A Monte Carlo Code to Simulate Extensive Air Showers}, Karlsruhe, Germany: Forschungszentrum Karlsruhe Report FZKA 6019, 1998.

\bibitem{PierreAuger}
The Pierre Auger Collaboration, \emph{Correlation of the Highest-Energy Cosmic Rays with Nearby Extragalactic Objects}, Science \textbf{318} (5852), 938 (2007)

\bibitem{4}
H.~Asorey \textit{et al.}, \emph{The LAGO space weather program: Directional geomagnetic effects, background
fluence calculations and multi--spectral data analysis}, in The $34^{th}$ International Cosmic Ray Conference, vol. PoS(ICRC2015), 142 (2015)

\bibitem{5}
I.~Sidelnik, \emph{The sites of the Latin American giant observatory}, in The $34^{th}$ International Cosmic Ray Conference, vol. PoS(ICRC2015), 665 (2015)

\bibitem{3}
H.~Asorey \textit{et al.}, \emph{LAGO: the Latin American Giant Observatory}, in The $34^{th}$ International Cosmic Ray Conference, vol. PoS(ICRC2015), 247 (2015)

\bibitem{Desorgher2004}
L.~Desorgher, \emph{MAGNETOCOSMICS: Geant4 application for simulating the propagation of cosmic rays through the Earth's magnetosphere}, available at \texttt{http://reat.space.qinetiq.com/septimess/magcos/}, 2004.

\bibitem{AgostinelliEtal2003}
S.~Agostinelli \textit{et al.}, \emph{Geant4 -- A Simulation Toolkit}, Nuclear Instruments and Methods A \textbf{506}, 250 (2003)

\bibitem{KumarTripathi2008}
R.~Kumar and A.~Tripathi, \emph{Root: A data analysis and data mining tool from cern}, Casualty Actuarial
Society E-Forum, 1 (2008).

\bibitem{HuedoEtal2001}
E.~Huedo \textit{et al}., \emph{The gridway framework for adaptive scheduling and execution on grids}, Scalable
Computing: Practice and Experience, \textbf{6} (3), 1, 2001.

\bibitem{BarberaFargettaRotondo2011}
R.~Barbera, M.~Fargetta and R.~Rotondo, \emph{A Simplified Access to Grid Resources by Science
Gateways}, In The International Symposium on Grids and Clouds and the Open Grid Forum, Taipei,
Taiwan, March 2011.

\bibitem{Montera}
M.~Rodr\'{\i}guez-Pascual \textit{et al.}, \emph{Montera: a framework for efficient execution of Monte Carlo codes on Grid infrastructures}, Computing and Informatics \textbf{32}, 113 (2013)

\bibitem{GWpilot}
A.J.~Rubio-Montero \textit{et al.}, \em`h{GWpilot: Enabling multi-level scheduling in distributed infrastructures with GridWay and pilot jobs}, Future Generation Computer Systems \textbf{45}, 25 (2015)

\bibitem{Gwcloud} 
A.J.~Rubio-Montero \textit{et al.}, \emph{User-Guided Provisioning in Federated Clouds for Distributed Calculations}, LNCS \textbf{9438}, 60 (2015)

\bibitem{TorresEtAl2011}
L.A.~Torres \textit{et al}., \emph{Implementación de un repositorio de datos científicos usando dspace}, E-Colabora, \textbf{1} (2), 101 (2011)

\bibitem{CamachoEtal2009}
R.~Camacho \textit{et al}., \emph{LAGOVirtual: A collaborative environment for the large aperture grb
observatory}. In R. Mayo, H. Hoeger, L. Ciuffo, R. Barbera, I. Dutra, P. Gavillet, and B. Marechal,
editors, Proceedings of the Second EELA2 Conferencem Choroní Venezuela, Madrid España, 2009.
EELA2, CIEMAT.

\bibitem{LewisDeCastroJones2012}
S.~Lewis \textit{et al.}, \emph{SWORD: Facilitating Deposit Scenarios}, D-Lib Magazine, \textbf{18} (1-2), (2012)

\bibitem{DART}
H.~Asorey \textit{et al.}, \emph{Data Accessibility, Reproducibility and Trustworthiness with LAGO Data Repository}, in The $34^{th}$ International Cosmic Ray Conference, vol. PoS(ICRC2015), 672 (2015)

\bibitem{CSE}
M.~Rodr\'{\i}guez-Pascual \textit{et al.}, \emph{A resilient methodology for accessing and exploiting data and scientific codes on distributed environments}, in Procs. 2015 IEEE $18th$ International Conference on Computational Science and Engineering, 319 (2015)

\end{thebibliography}
\end{document}